\documentclass[a4paper]{jpconf}
\usepackage{graphicx}
\usepackage{iopams}

\begin{document}

\title{Method of EAS`s Cherenkov and fluorescent light separation using silicon photomultipliers}

\author{Dmitry Chernov$^{1,*}$,
        Elena Bonvech$^1$,
        Timur Dzhatdoev$^1$,
        Miroslav Finger$^{3,4}$,
        Michael Finger$^{3,4}$
        Vladimir Galkin$^{1,2}$
        Gali Garipov$^1$,
        Vladimir Kozhin$^1$,     
        Dmitry Podgrudkov$^{1,2}$ and
        Alexander Skurikhin$^1$}

\address{$^1$  Skobeltsyn Institute of Nuclear Physics, M.V.Lomonosov Moscow State University, 1(2), Leninskie gory, GSP-1, Moscow, 119991, Russian Federation}
\address{$^2$ Department of Physics,  M.V.Lomonosov Moscow State University, 1(2), Leninskie gory, GSP-1, Moscow, 119991, Russian Federation}
\address{$^3$ Faculty of Mathematics and Physics, Charles University, Ke Karlovu 3, 121 16, Prague 2, Czech Republic}
\address{$^4$ Joint Institute for Nuclear Research, Joliot-Curie, 6, Dubna, Moscow region, 141980, Russian Federation}

\ead{$^{*}$chr@dec1.sinp.msu.ru}

\begin{abstract}
Preliminary results on the development of a separation method for Cerenkov (CL) and fluorescence (FL) light from EAS are shown. The results are based on the measurement of attenuation coefficients of CL and FL for different filters. A total of six optical filters were investigated: filters from optical glass UFS-1, UFS-5, FS6 (analogue BG3) and interference filters SL~360-50, SL~280-380, FF01-375/110. The measurements were performed using silicon photomultipliers (SiPM). To improve existing fluorescent light detectors, a segment of 7 SiPM was developed, which would be able to separate both components of the light flux from EAS at the level of primary data processing.
\end{abstract}

\section{Introduction}
Direct and scattered Cherenkov light (CL) is one of the relevant contributions to the uncertainty of the measured flux of fluorescent light (FL)  from extensive air showers (EAS). The impact of backscattered CL was noted when modelling the response of several detectors~\cite{EUSO2015},\cite{Auger2010},\cite{ Unger2008}. The problem of reliable CL and FL separation is relevant for a better detection of high energy EAS and estimation of the primary particle parameters. This study is the first step in search of both theoretical approaches to the problem and possible design of electronics for actual and future detectors.

\section{The method of separation FL and CL}
A separation method based on the simultaneous recording of light from one `optical pixel’ by two or more pairs of silicon photomultipliers (SiPMs) is proposed. The first SiPM detects the incoming light flux in the wavelength band of its maximal sensitivity. The second SiPM detects it through an optical ultraviolet (UV) filter for fluorescent light (FL) separation. If the SiPMs' sensitivity characteristics, absorption characteristics of the filter elements and the spectra of fluorescent~\cite{AIRFLY2016} and Cherenkov light are known, one can calculate the contribution of each component to the total light flux. Upon the completion of this work it will be possible to separate fluorescent and Cherenkov light at the stage of on-board primary processing of recorded data. This method will increase the methodological accuracy of fluorescent light measurements.

\begin{figure}
\begin{center}
\includegraphics[width=0.8\textwidth]{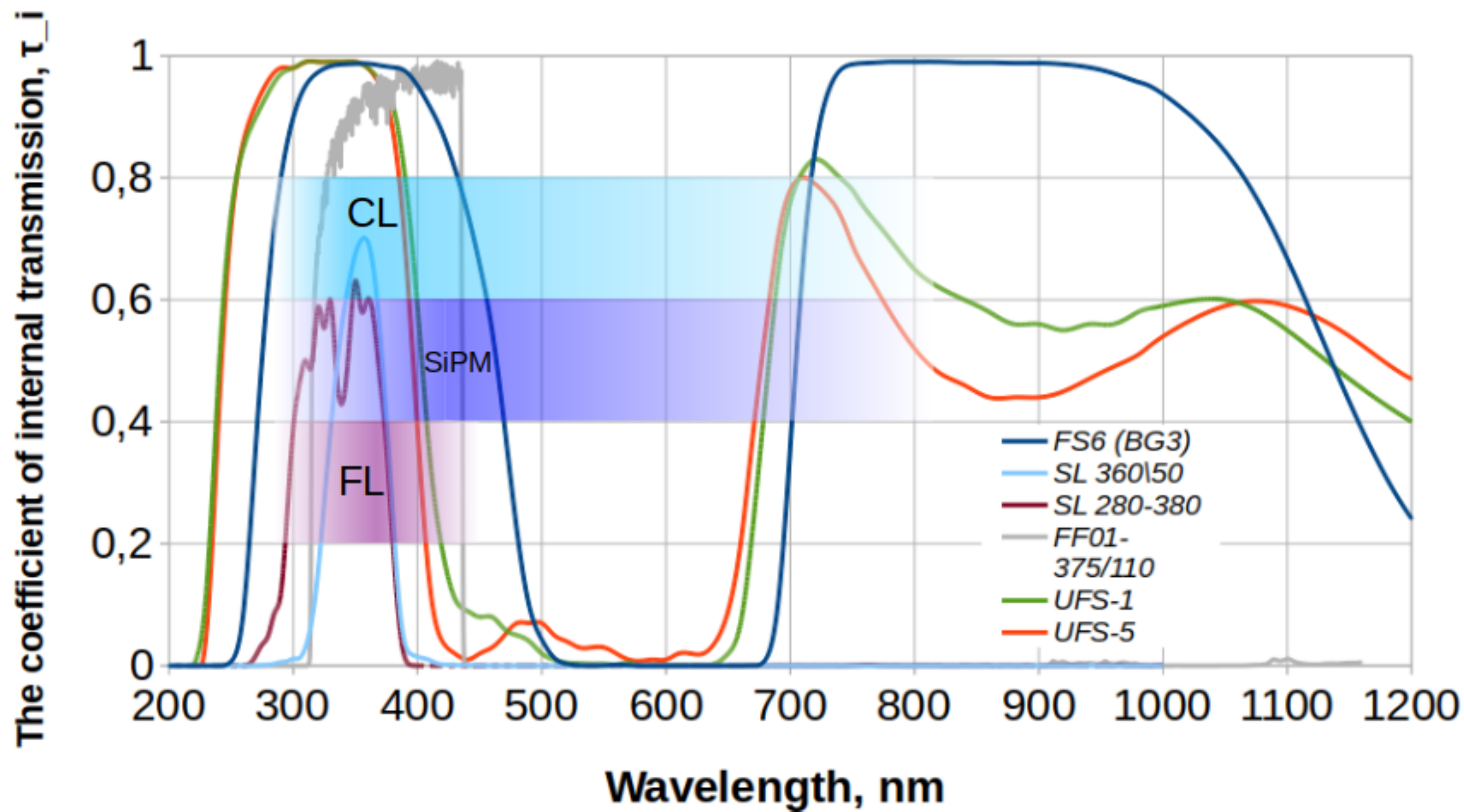}
\end{center}
\caption{Spectral characteristics of light filters.}
\label{Spect_filters} 
\end{figure}

In this work both ordinary colored-glass filters and interference filters were studied. In theory the interference filters should allow a better FL separation due to their narrow pass-through band in the UV region. The spectral characteristics of all studied filters are show on fig.~\ref{Spect_filters}. 

\begin{figure}
\begin{center}
\includegraphics[width=0.8\textwidth]{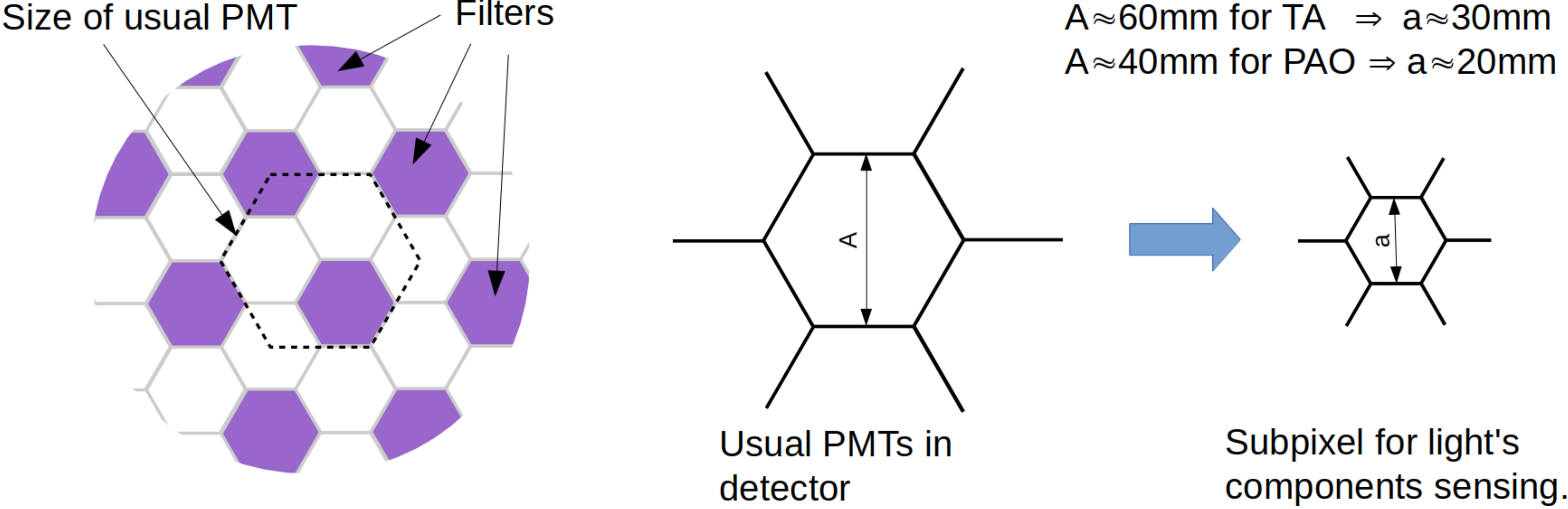}
\end{center}
\caption{PMT mosaic modification example. The areas ratio (with filter)/(without filter) is equal to 1/2.}
\label{method}
\end{figure}

In fig.~\ref{method} the possible modification of the common photomultiplier tubes (PMT) mosaic is shown on the example of the Telescope Array (TA)~\cite{TA2012} and Pierre Auger Observatory~\cite{Auger2015} detectors. In this case the signal loss on the filters is minimal since only one third of the mosaic area is covered by the filters. However, it is worth noting that the proposed method is effective only when the light spot from the remote point source on the mosaic is two or more times larger than the SiPM mosaic subpixel size.    

\section{FL and CL test stands}

\begin{figure}
\begin{center}
\includegraphics[width=1.0\textwidth]{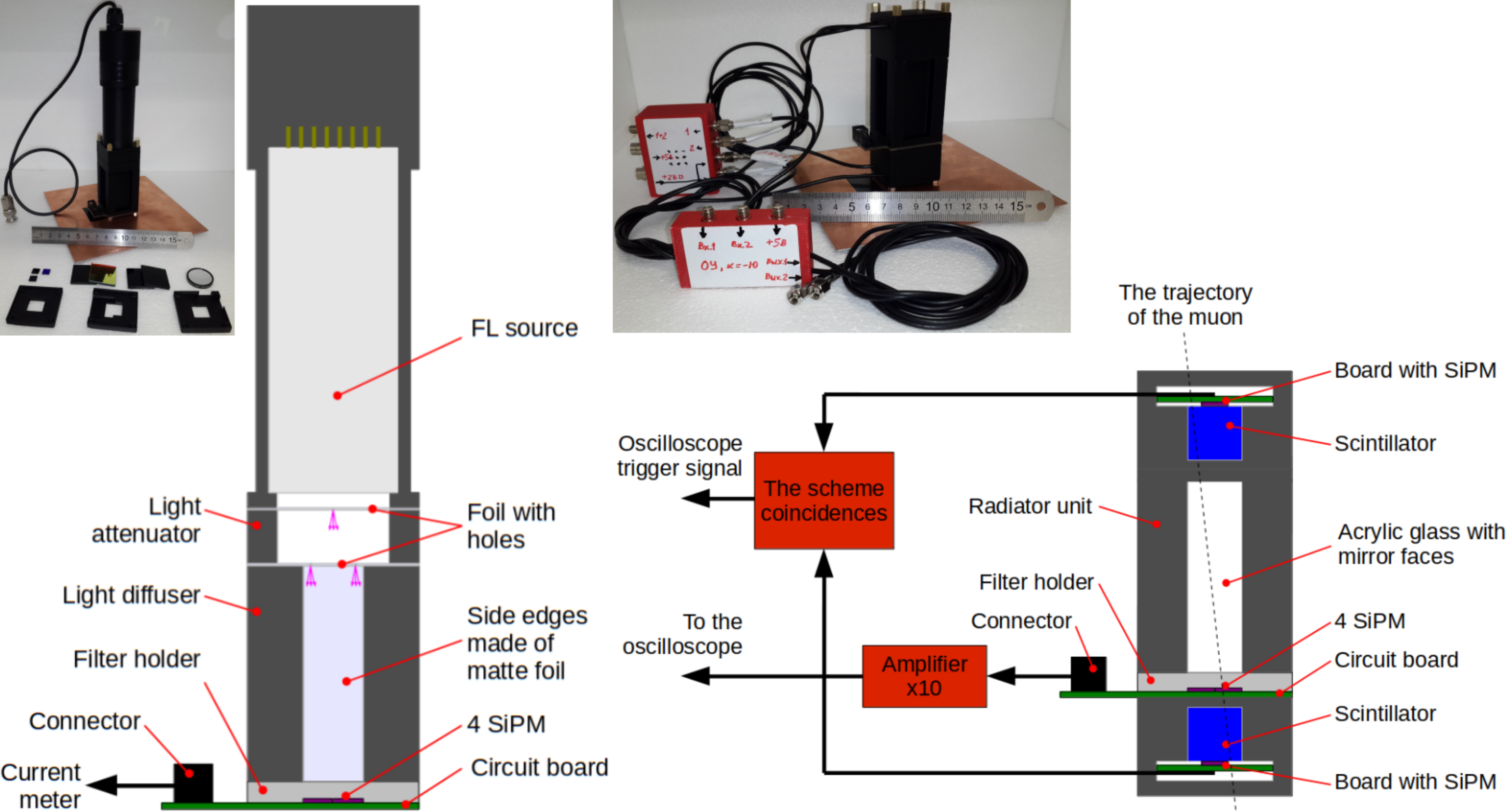}
\end{center}
\caption{\label{FL_CL_stands}Schemes and photos of the stands for filter (UFS-1, UFS-5, FS6, SL 360$\setminus$50, SL 280-380, FF01-375/110) attenuation coefficient measurements. The FL stand is on the left and the CL one is on the right.}
\end{figure}

On the right side in fig.~\ref{FL_CL_stands} are the general scheme and photo of the stand for the optical filters testing with CL. CL photons are generated by cosmic ray muons passing through a $15\times15\times50$~mm acrylic radiator (acrylic glass without a UV stabilizer) with 5 reflective sides (the bottom side was transparent). The radiator's bottom side was matted to create an isotropic CL flux. The selection of vertical muons passing through the whole radiator is done using two $15\times15\times15$~mm scintillator blocks with a SiPM (shown in blue on the scheme) placed above and below the radiator. Signals from the SiPMs are routed to the coincidence scheme that in turn triggers the recording on a digital oscilloscope. The tested filters are installed between the radiator and 4 measuring SiPMs using different fittings. Each muon passing through the radiator gives around 1800 CL photons, but due to scattering, multiple reflections and bottom side absorption only about 25\% of photons reach the measuring SiPMs. For more reliable measurements the SiPMs were connected in parallel in diagonal pairs. The resulting 2 signals were amplified by fast operational amplifiers~AD8011 and then recorded by the oscilloscope. This scheme allows to have an average pulse amplitude of 42~mV from each channel with noise level less than 5~mV.

On the left side in fig.~\ref{FL_CL_stands} are the general scheme and photo of the stand for testing optical filters with FL. The FL source was manufactured~\cite{MELZ_PMT} on request and is described in detail in~\cite{FL_CL_2017}. The stand consists of a FL source, light flux attenuator, a diffuser, filter housing and~4~SiPMs. The light attenuator consists of two parallel foils with multiple pinholes in them situated approximately 12~mm apart of each other. The pinholes are arranged so that direct light from the FL source does not reach the filters and SiPMs. The diffuser consists of a $15\times15\times50$~mm light guide made of matted foil to match the measuring geometry of the CL test stand. The optical filters, their housing and measurement SiPM arrangement and circuitry was the same as on the CL stand described above. The only difference is that FL is continuous. In this case it is more convenient to measure and compare the SiPM anode currents. For the SiPM linearity test two equal diodes with 403~nm main wavelength were used. The diodes illuminated SiPMs one by one with 5 different voltages and in combinations. The SiPMs proved to have high linearity.

\section{Preliminary results}
Using the stands described above a series of tests were carried out to determine the attenuation coefficients for 6 filters (see fig.~\ref{Spect_filters}) for both CL and FL. The preliminary results are shown in table~\ref{table}. 

\begin{table}[t]
\caption{Preliminary results on the transparency coefficient measurements for FL and CL of conventional optical filters (UFS-1, UFS-5, FS6) and interference filters (SL 360$\setminus$50, SL 280-380, FF01-375/110).}
\label{table}
\begin{center}
\begin{tabular}{lccc}
\br
Filter &FL &CL &Arbitrary separation efficiency\\
\mr
FF01-375/110        & $0.73\pm 0.07$ & $0.39\pm0.15$ & 1.87\\
UFS-5               & $0.65\pm 0.01$ & $0.36\pm0.26$ & 1.81\\
UFS-1               & $0.71\pm 0.01$ & $0.48\pm0.25$ & 1.48\\
FS6 (BG3)           & $0.81\pm 0.01$ & $0.55\pm0.18$ & 1.47\\
SL 280-380          & $0.36\pm 0.01$ & $0.33\pm0.11$ & 1.09\\
SL 360$\setminus$50 & $0.28\pm 0.01$ & $0.28\pm0.09$ & 1.00\\
\br
\end{tabular}
\end{center}
\end{table}

For filter comparison the `Arbitrary separation efficiency' characteristic, defined as the ratio of FL transparency to CL transparency, was introduced. The best characteristics has the interference filter~FF01-375/110 and colored glass filter UFS-5.The FF01-375/110 filter has a good FL transparency but as a drawback has a high cost. Moreover, the filters' angular properties need more careful analysis since in many actual experiments the light falls on the sensitive part of the detector in a wide range of angles while in this study it was collected in a narrow angle. 

Since the filters were installed over SiPMs without an optical contact the transparency coefficients in table~\ref{table} also include reflection losses on the SiPM surface. Since FF01-375/110 compared to other filters has a `perfectly' polished surface that gives visible flares the measured transparency coefficient has a higher uncertainty. In the next planned measurement series this will be corrected using a different diffuser design. 

A significant uncertainty in CL measurements comes from two major sources: 
\begin{enumerate}
\item muon trajectory in the radiator and 
\item low muon flux --- $\sim$3~muons per hour. 
\end{enumerate}
The first factor gives a difference factor of up to 2 in two measuring channels, but does not change the average value. The second factor does not allow to gather enough statistics in a reasonable time to reduce the statistical error. While the expositions can be increased by 10 times it still will not be nearly enough to reach the precision of FL measurements. To solve this issue the measurement of CL filter attenuation coefficients is planned to be carried out using CL from EAS. The measurements are planned to take place on a test site with low light pollution and aerosol levels using a 0.3~m$^2$ 470~mm curvature radius mirror and a set of 7 SiPMs with light collectors. In comparison with the stand described above the new setup will increase the light flux per SiPM by more than 15 times and in total increase to about 150 events per hour.

\section{Development of the detector}

The detector is designed for ground and space based optical experiments for high and ultrahigh energy cosmic ray studies. 

\begin{figure}
\begin{center}
\includegraphics[width=0.9\textwidth]{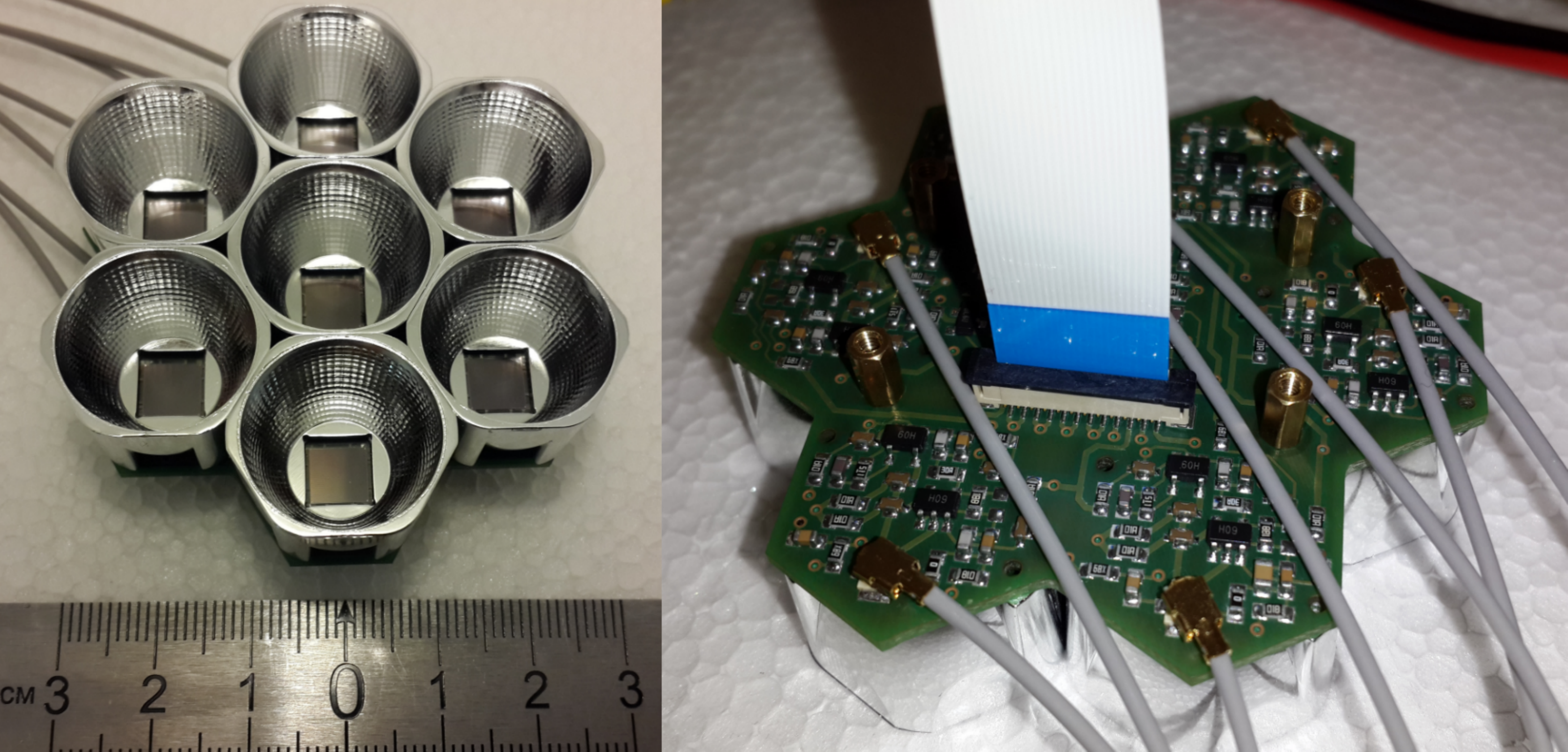}
\end{center}
\caption{\label{Seven_SiPM} Matrix of seven SiPM SensL MicroFC-60035-SMT with light collectors and amplifiers.}
\end{figure}

It is planned to create a sensitive module utilizing light collectors. The module will consist of a board with 7 SiPMs \cite{SensL} and a set of light collectors (fig.~\ref{Seven_SiPM} right and left respectively), amplifiers and set of additional sensors (temperature, pressure etc). The light collectors effectively enlarge the pixel area to $\sim$3~cm$^2$. Unlike the commercially available ones, the new matrix allows to set the supply voltage individually for each SiPM for sensitivity equalization across the module (and between the modules). It is planned that the whole sensitive matrix will consist of 7 modules , i.e. 49 pixels in total. The effective sensitive area should be 4--5 times larger than for a single module (due to the mirror attenuation by the matrix).

\begin{figure}
\begin{center}
\includegraphics[width=0.9\textwidth]{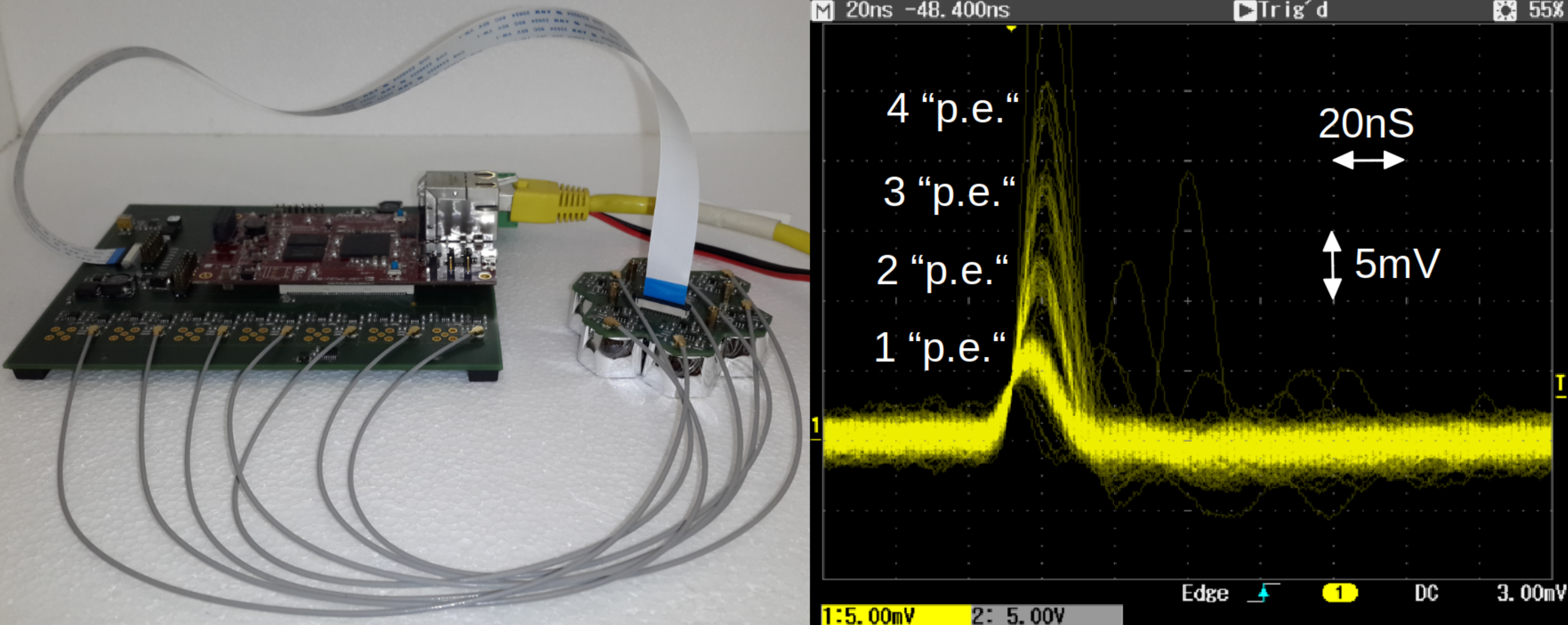}
\end{center}
\caption{\label{Electronic_signal} Left panel: an electronic board for counting photoelectrons with a Zynq-7020 FPGA debugger. Right panel: An amplified signal of SiPM's fast output.}
\end{figure}

On fig.~\ref{Electronic_signal} the 7-channel photoelectron counting board prototype is shown. The board houses a debug module with a FPGA chip and receives commands via ethernet interface. It also houses the SiPM power source ($-24$ \ldots$-29$~V), digital-to-analogue converters (DAC) for individual SiPM sensitivity control, comparators for photoelectron counting, DACs for the their level selection and DACs for sensor readings. The 7-SiPM module is connected to the board. The white cable is used for module powering and 7 gray cables (micro-coaxial cables) connect the SiPM amplifiers to the comparators.

\section{Conclusion}
A method for separating Cherenkov and the fluorescent light from EAS is being developed. Preliminary measurement results indicate good prospects from the application of the proposed method. The first elements of the detector using silicon photomultipliers for the realization of this method are produced.

\ack
This work was funded by Russian Foundation for Basic Research (RFBR) project No.16-02-00777.


\section*{References}
\begin{thebibliography}{9}

\bibitem{EUSO2015}
  Haungs A et al. for JEM-EUSO Collaboration 
  2015
  {\it J. Phys.: Conf. Series} 
  {\bf 632}  012092

\bibitem{Auger2010}
  The Pierre Auger Collaboration
  2010
  {\it Astropar. Phys.} 
  {\bf  33} 108
  
\bibitem{Unger2008}
  Unger M,  Dawson B R,  Engel R, Schussler F, Ulrich R
  2008
  {\it Nucl. Instrum. Methods} 
  {\bf A588}  433

\bibitem{AIRFLY2016}
  Ave M et al. for AIRFLY Collaboration
  2007
  {Measurement of the pressure dependence of air fluorescence emission induced by electrons},
  {\it Astropart. Phys.} 
  {\bf 28}  41
  
\bibitem{TA2012}
  The Telescope Array Collaboration
  2012
  {\it Nuclear Instruments and Methods in Physics Research A}
  {\bf 676} 54–65
  
 \bibitem{Auger2015}
  The Pierre Auger Collaboration
  2015
  {\it Nuclear Instruments and Methods in Physics Research A}
  {\bf 798} 172-213

\bibitem{MELZ_PMT}
  MELZ FEU Ltd
  {\it http://www.melz-feu.ru/}
  
\bibitem{FL_CL_2017}
  Chernov D V, Bonvech E A, Dzhatdoev T A, Finger Mir., Finger Mich., Galkin V I, Garipov G K, Kozhin V A, Podgrudkov D A, Skurihin A V
  2017
  {Position-sensitive SiPM detector for separation of Cherenkov and fluorescent light of EAS},
  {\it PoS} 
  ICRC(2017)444  
 
\bibitem{SensL}
  SensL Technologies Ltd 
  {\it http://sensl.com/}
  
\end{thebibliography}
\end{document}